\documentstyle[rmp,aps]{revtex}

\begin{document}
\title{Where the Tsallis Statistic is valid? }
\author{L.Vel\'{a}zquez $^{1}$, F.Guzm\'{a}n $^{2}$}
\address{$^{1}$Departamento de Fisica, Universidad de Pinar del Rio, Cuba\\
e-mail: luisberis@geo.upr.edu.cu \\
$^{2}$Departamento de Fisica Nuclear, Instituto Superior de Ciencias y\\
Tecnolog\'{i}as Nucleares, Ciudad Habana, Cuba.\\
e-mail: guzman@info.isctn.edu.cu }
\date{\today}
\maketitle

\begin{abstract}
In the present paper are analyzed the conditions for the validity of the
Tsallis Statistics. The same have been done following the analogy with the
traditional case: starting from the microcanonical description of the
systems and analysing the scaling properties of the fundamental macroscopic
observables in the Thermodynamic Limit. It is shown that the Generalized
Legendre Formalism in the Tsallis Statistic only could be applied for one
special class of the bordering systems, those with non exponential growth of
the accessible states density in the thermodynamic limit and zero-order
divergency behavior for the fundamental macroscopic observables, systems
located in the chaos threshold.
\end{abstract}

\section{Introduction}

In the last years many researchers have been working in the justification of
the Tsallis Formalism, but many of them have pretend to do it in the context
of the Information Theory$\left[ 1,2,3\right] $ without appeal to the
microscopic properties of the systems. Our opinion is that all of these
attempts lead to the conclusion of the {\it non uniqueness} of the entropy
concept by mean of {\it probabilistic interpretation}. That is why we
consider that the statistical description of nonextensive systems has to
start from the microscopic characteristics of them.

Boltzmann$\left[ 4\right] $, Gibbs$\left[ 5\right] $, Einstein$\left[ 6,7%
\right] $ and Ehrenfests$\left[ 8\right] $ recognized the hierarchical
primacy of the microcanonical ensemble with regard to the others statistical
ensembles. These last ones can be derived starting from the first by the
consideration of certain particular conditions: {\it the extensivity
postulates}. This is the essential ideas to generalize the traditional
results for the justification of the Tsallis Statistics, to put it on the
level of the microscopic description, in the ground of the Mechanics.

With these ideas A. Plastino and A. R. Plastino$\left[ 9\right] $ have
proposed one way to justify the q-generalized canonical ensemble with
similar arguments employed by Gibbs himself in deriving his canonical
ensemble. It is based in the consideration of a closed system composed by a
subsystem with a weakly interaction with a {\it finite thermal bath}. They
showed that the macroscopic characteristics of the subsystem are described
by a Tsallis potential distribution, relating the {\it q entropy index} with
the finiteness of the last one.

Another attempt was made by S. Abe and A. K. Rajagopal$\left[ 10,11\right] $%
, with the same idea: a closed system composed by a subsystem with a weakly
interaction with a very large thermal bath, but this time analyzing the
behavior of the systems around the equilibrium, considering the last one as
the most probable configurations. He shown that the Tsallis generalization
of the Boltzmann-Gibbs Distribution can be obtained if the counting rule is
modified, introducing the Tsallis generalization of the logarithmic function
for arbitrary entropic index.

However we disagree with these interpretations. So far it has been said that
the statistic of Tsallis allows to extend the study for systems that are
anomalous for the traditional thermodynamic, systems with of long-range
correlations, due to the presence of long-range interactions with a dynamics
of non markovianity stochastic processes, where the {\it entropic index
gives a measure of the non extensivity of a system, an intrinsic
characteristic of the same}$\left[ 12\right] $. That is the reason why we
find the identification of this parameter with the finiteness of a thermal
bath an artificial, outside of the supposed application context that has the
Statistic of Tsallis.

For example, for the Plastino and Plastino analysis, What is that we
consider by a thermal bath in astrophysical systems or a atomic nucleus?
What is that we consider by a finite thermal bath in the turbulent fluids or
a non screened plasma?

By other hand, for the Abe and Rajogopal analysis, they insisted that {\it %
there is an arbitrariness in selection of the counting rule} that which
determinate the form of the distribution. Although the selection of the
counting rule do not have so much influence for configurations of the system
near from the equilibrium, for configurations far from the equilibrium this
selection is very important since this fact mark the difference among all
the generalized canonical ensembles. In their works they do not establish a
criterium that allows to define in univocal manner the selection of the
counting rule.

However, this is not the case of the Boltzmann-Gibbs Statistics, because its
counting rule is supported by the scaling behavior of the states density 
{\it in the Thermodynamic Limit }(ThL), by mean of the {\it scaling \
behavior} of the fundamental macroscopic observables with the increasing of
the degree of freedom of the system, and its Thermodynamic Formalism, based
on the {\it Legendre Transformation} between the Thermodynamic potentials,
by {\it equivalence between the microcanonical and the canonical ensembles
in the ThL} too .

At the present work we pretend to analyze the conditions to satisfy by the
system in order to justify the application of the Tsallis Statistics in
them. This time, we will pay special attention to equivalency of the
ensembles in the ThL

\section{Microcanonical ensemble: its geometrical aspects and scaling
properties.}

It is very well know that the microcanonical ensemble are equivalent to the
traditional canonical and Gran-canonical ensembles in the ThL of the
infinitely many particles interacting by short range interactions if the
system is homogeneous.

However, in the nature we find examples of systems where the thermodynamic
limit does not take place, since they are not composed by a huge number of
particles. We find examples of these in the molecular and atomic clusters. A
very interesting problem is the extension of the macroscopic description for
this kind of systems as well as the question about when could be considered
that these systems are found in the thermodynamic limit.

An important step toward the resolution of these problems was accomplished
by D.H.E. Gross with the development of the {\it Microcanonical
Thermostatistics }$\left[ 13,14\right] $, formalism based on the
consideration of the microcanonical ensemble. In this approach is possible
to accomplish the description of some finite systems, extending thus the
study of {\it phase transitions} in them. On the other hand, it is
attractive the possibility that has the same of exploring the behavior of
the macroscopic observables with the increasing of the number of particles,
until the convergency in some cases in an ordinary extensive system,
permitting thus to give a valuation about when the thermodynamic limit is
reached in the system.

In our opinion, this study reveled two very useful conclusions:

\begin{itemize}
\item  Any closed system could be exhaustively described by the
microcanonical ensemble without matter how many particles it has. In fact,
this is the only way to generalize the equilibrium thermodynamics for
arbitrary closed system without appealing to anything outside of the
mechanics.

\item  It is possible that in the thermodynamical limit the microcanonical
ensemble could be {\it equivalent} to some \ \ \ \ \ \ \ \ \ \ \ \ \ \ \ \ \
\ \ \ \ \ generalized \ canonical ensemble.
\end{itemize}

Following the analogy with the traditional case, it would be possible to
associate to this generalized canonical ensemble a probabilistic
interpretation of the entropy concept with the same style of Shannon
-Boltzmann-Gibbs entropy and the non extensive entropy of \ Tsallis, being
this entropy equivalent with some generalized definition of the Boltzmann
entropy in the thermodynamic limit. In the last case, \ it is really
interesting point because open the door to a possibility to obtain the
entropic index q of the Tsallis formalism without the necessity of invoke to
additional postulates.

A common denominator of all \ {\it Probabilistic \ Thermodynamics \
Formalisms} (PThF), \ those \ derived from a probabilistic interpretation of
the entropy concept, is the description of the equilibrium macroscopic state
of the systems by means of the intensive parameters of the generalized
canonical distributions, $\beta $, that is to say, they are supported by the
validity of the{\it \ zero principle of the Thermodynamics}.

In our previous works $\left[ 15,16\right] $ we centered in the geometrical
aspects that posses the {\it Probabilistic Distribution Functions} (PDF) of
the different ensembles. In the case of the microcanonical ensemble we
showed that its PDF is invariant under the transformation group of {\it %
local reparametrizations or Diffeomorfism} of the space of the integrals of
movement, $Diff\left( \Im _{N}\right) $, being this the maximal symmetry
group that a geometrical theory could have, being this associated with the
local properties of a general space. By other hand, all the PDF derived from
the PThF must depend on the integrals of movement in a lineal combination
with the canonical intensive parameters: 
\begin{equation}
p\left( X;\beta ,N\right) =F(\beta ,N;\beta \cdot I_{N}\left( X\right) )
\end{equation}
therefore, the most general symmetry group that preserve this functional
form is the {\it general lineal group}, $GL\left( R^{n}\right) $(n is the
dimension of $\Im _{N}$), which is related with the euclidean vectorial
spaces. It is supposed that a {\it spontaneous symmetry breaking} happens
during the thermodynamic limit, from $Diff\left( \Im _{N}\right) $ to $%
GL\left( \text{\ss }^{n}\right) $, where \ss $^{n}$ is the space of the
generalized canonical parameters. The way in that this hypothetical symmetry
breaking happens will determine the specific form of the statistics in the
ThL. 

What we understand by a spontaneous symmetry breaking? Ordinarily, the
microcanonical ensemble could be described equivalently through any
representation of the abstract space of the integrals of movement of the
distribution. However, with the increasing of the degree of freedom of the
system, some specific representation will be more addecuate to describe the
macroscopic state because they reflect in better manner the general
properties of the system: i.e. in the case of the traditional systems, {\it %
the extensivity}.

In complete analogy with the traditional analysis, we identify the cause of
this spontaneous symmetry breaking with the scaling properties of the
fundamental macroscopic observables , the behavior of the integrals of
movement and the accessible states density with the increasing of the degree
of freedom of the system.

In general, the scaling properties of the systems in the ThL could be
diverse. In order to be more specific,we will consider here the following
scaling behavior of the state density: 
\begin{eqnarray}
a)\text{ }W\left( I,N\right)  &\simeq &\exp \left( \gamma \left( I,N\right)
N^{\tau }\right) \text{ \ \ }\ b)\text{ }W\left( I,N\right) \simeq \gamma
\left( I,N\right) N^{\tau }\text{ } \\
\text{\ \ }c)\text{ }W\left( I,N\right)  &\simeq &\gamma \left( I,N\right)
\ln ^{\tau }N\text{ \ \ \ \ \ \ \ \ }d)\text{ }W\left( I,N\right) \simeq
N^{\gamma \left( I,N\right) }\text{ }  \nonumber
\end{eqnarray}
where $\gamma \left( I,N\right) $ is an scaling invariant function of the
integrals of movement, and $\tau $ is an scaling invariant parameter.
Extensivity is associated with the dependency a) with $\tau =1$ , when the
interacting forces have a short-range and there is independency between
different parts of the system. Any deviation of this behavior is a sinal of
non-extensivity, the presence of long-range correlation in the system. The
systems with scaling behavior given by $\left( 2.a\right) $ with $0<\tau <1$
we will refer as {\it pseudoextensive systems} and the others by {\it %
bordering systems}. The reasons for the last denominations will be
understand later.

In order to aim this discussion in the analysis of the conditions for the
validity of the Tsallis formalism, we will fit all above behaviors by:

\begin{equation}
W\left( I,N\right) \simeq e_{q}\left[ \gamma \left( I,N\right) \ln
_{p}\left( N\right) \right]
\end{equation}
where $e_{q}\left( x\right) $ and $\ln _{q}\left( x\right) $ are the
generalized q-exponential and q-logarithmic function of the Tsallis
formalism:

\begin{equation}
e_{q}\left( x\right) =\left[ 1+\left( 1-q\right) x\right] ^{\frac{1}{1-q}}%
\text{ \ \ \ }\ln _{q}\left( x\right) =\frac{x^{1-q}-1}{1-q}\equiv
e_{q}^{-1}\left( x\right)
\end{equation}

In this case, it is convenient define the entropy by:

\begin{equation}
S_{B-T}=\ln _{q}\left[ W\right]
\end{equation}
in order to access to the invariant scaling function $\gamma \left(
I,N\right) $, because it contains all the useful information that we can
extract from the macroscopic description of the system. We recognize
immediately the Tsallis generalization of the Boltzmann entropy.

\section{The Legendre formalism}

In this section we will pay attention to the relationship between the
microcanonical and the canonical ensemble. To do this we will reproduce
first the general analysis realized by D.H.E.Gross in deriving his
microcanonical thermostatistics. After, in analogy with the previous
exposition, we will consider the q-generalization of the canonical ensemble
of Tsallis.

\subsection{The case of pseudoextensive systems.}

For this kind of system we will consider the usual definition of Boltzmann
entropy with $q=1$:

\begin{equation}
S_{B}=\ln \left[ W\right]
\end{equation}
In this case, the canonical PDF will be given by:

\begin{equation}
\omega \left( X;\beta ,N,a\right) =\frac{1}{Z\left( \beta ,N,a\right) }\exp %
\left[ -\beta \cdot I_{N}\left( X;a\right) \right]
\end{equation}
where $Z\left( \beta ,a,N\right) $ is the partition function:

\begin{equation}
Z\left( \beta ,N,a\right) =\int \exp \left[ -\beta \cdot I_{N}\left(
X;a\right) \right] dX
\end{equation}
Introducing the delta function in the above integral, it can be rewrite as:

\begin{equation}
Z\left( \beta ,N,a\right) =\int \exp \left[ -\beta \cdot I\right] dI\int
\delta \left[ I-I_{N}\left( X;a\right) \right] dX
\end{equation}
The integration of the delta function yields the density of states:

\begin{equation}
W\left( I,N,a\right) =\Omega \left( I,N,a\right) \delta I_{o}=\left( \int
\delta \left[ I-I_{N}\left( X;a\right) \right] dX\right) \delta I_{o}
\end{equation}
where $\delta I$ is a {\it suitable} constant volume element to make $W$
dimensionless. We arrive finally to the {\it Laplace Transformation }of the
states density::

\begin{equation}
Z\left( \beta ,N,a\right) =\int \exp \left[ -\beta \cdot I\right] W\left(
I,N,a\right) \frac{dI}{\delta I_{o}}
\end{equation}

The Laplace Transformation establishes the connection between the Boltzmann
entropy with the fundamental thermodynamics potential of the canonical
distribution, {\it the Planck Potential}:

\begin{equation}
P\left( \beta ,N,a\right) =-\ln Z\left( \beta ,N,a\right)
\end{equation}

We adopted the Planck Potential because it do not introduce any preference
with an specific integral of movement of the microcanonical distribution, in
order to be consequent with the $GL(R^{n})$ invariance. The last integral
can be rewrite as:

\begin{equation}
\exp \left[ -P\left( \beta ,N,a\right) \right] =\int \exp \left[ -\beta
\cdot I+S_{B}\left( I,N,a\right) \right] \frac{dI}{\delta I_{o}}
\end{equation}

Immediately, we recognized in the exponential argument the {\it Legendre
Transformation} between the thermodynamic potentials. If the integrals of
movements and the Boltzmann entropy have the same scaling behavior in the
ThL, this integral will have a very sharp peak around the maximum value{\it %
\ if this maximum exists} and therefore, the main contribution to this
integral will come from this maximum.

If we define $\widetilde{P}\left( \beta ,N,a\right) $ by:

\begin{equation}
\widetilde{P}\left( \beta ,N,a\right) =Max_{I=I_{M}}\left[ \beta \cdot
I-S_{B}\left( I,N,a\right) \right]
\end{equation}
The maximization yields:

\begin{equation}
\beta _{\mu }=\left. \frac{\partial }{\partial I^{\mu }}S_{B}\right|
_{I=I_{M}}
\end{equation}
being this the usual relation between the canonical intensive parameters and
the entropy. Introducing the {\it curvature tensor}:

\begin{equation}
\left( K_{B}\right) _{\mu \nu }=\left. \frac{\partial }{\partial I^{\mu }}%
\frac{\partial }{\partial I^{\nu }}S_{B}\right| _{I=I_{M}}
\end{equation}
the Laplace transformation will be estimated by:

\begin{eqnarray}
\exp \left[ -P\left( \beta ,N,a\right) \right] &\simeq &\int \exp \left[ -%
\widetilde{P}\left( \beta ,N,a\right) \right] \exp \left[ -\frac{1}{2}\left(
I-I_{M}\right) ^{\mu }\cdot \left( -K_{B}\right) _{\mu \nu }\cdot \left(
I-I_{M}\right) ^{\nu }\right] \frac{dI}{\delta I_{o}} \\
&\simeq &\exp \left[ -\widetilde{P}\left( \beta ,N,a\right) \right] \frac{1}{%
\delta I_{o}\det^{\frac{1}{2}}\left( -\frac{1}{2\pi }\left( K_{B}\right)
_{\mu \nu }\right) }
\end{eqnarray}

The above integral will be well defined if all the eingenvalues of the
curvature tensor are negatives, if the Boltzmann entropy if locally concave
around the maximum. In this case, in the canonical ensemble there will be
small fluctuations of the integrals of movement around its mean values with
a standard desviation:

\begin{equation}
\frac{\sqrt{\overline{\left( I^{\mu }-I_{M}^{\mu }\right) \left( I^{\nu
}-I_{M}^{\nu }\right) }}}{I_{M}^{\mu }I_{M}^{\nu }}\sim \frac{1}{N^{\frac{%
\tau }{2}}}
\end{equation}
Thus, the Planck potential could be obtained by mean of the Legendre
formalism:

\begin{equation}
P\left( \beta ,N,a\right) \simeq \beta \cdot I-S_{B}\left( I,N,a\right)
\end{equation}
and the Boltzmann entropy will be equivalent with the
Shannon-Boltzmann-Gibbs entropy:

\begin{equation}
S_{S-B-G}=-%
\mathrel{\mathop{\sum }\limits_{k}}%
p_{k}\ln p_{k}\simeq S_{B}
\end{equation}
However, if the concavity condition for the entropy is not hold, there will
be a catastrophe in the Laplace transformation and the canonical ensemble
will not be able to describe the system. Ordinarily, this situation is a
sinal of the occurrence of phase transitions in the system.

As it has been shown, although the pseudoextensive systems are nonextensive
in the usual sense, their study could be carry out with the same formalism
employing for the ordinary extensive systems with appropriate selection of
the representation of the space of integrals of movement. The Eq. $(15)$
leads to condition of the {\it homogeneous scaling of the integrals of
movement and the entropy} in order to satisfy the requirement of the scaling
invariance of the canonical parameters, the validity of the zero principle
of the thermodynamics. Here we have the cause of the spontaneous symmetry
breaking. Although in the microcanonical ensemble all the representation of
the integrals of movement are equivalent, in the generalized canonical
ensemble only are possible those representations with an {\it homogeneous
nondegenerated scaling}. To understand the term {\it nondegenerate }\ let us
show the following example. Let be $A$ and $B$ two integrals of movement
with different scaling behavior:

\begin{equation}
A\sim N^{a}\text{;\ }B\sim N^{b}\text{ with }a>b
\end{equation}
In another representation, these integrals could be represented equivalently
in the microcanonical ensemble by:

\begin{equation}
I^{\pm }=A\pm B
\end{equation}
but in this case the scaling behavior of them are:

\begin{equation}
I^{\pm }\approx A\sim N^{a}
\end{equation}

In the ThL $I^{\pm }$ have an homogeneous scaling but they are not
independent. The canonical parameters derived of $I^{\pm },$ $\beta ^{\pm }=%
\frac{\partial S_{q}}{\partial I^{\pm }}$, will be identically in the ThL.
This leads to the {\it trivial vanishing }of the {\it curvature tensor
determinant in ThL}:

\begin{equation}
\mathrel{\mathop{\lim }\limits_{N\rightarrow \infty }}%
N^{2a}\det \left( K_{\mu \nu }\right) =%
\mathrel{\mathop{\lim }\limits_{N\rightarrow \infty }}%
N^{2a}\det \left( 
\begin{tabular}{ll}
$\partial _{+}\partial _{+}S_{B}$ & $\partial _{-}\partial _{+}S_{B}$ \\ 
$\partial _{+}\partial _{-}S_{B}$ & $\partial _{-}\partial _{-}S_{B}$%
\end{tabular}
\right) \equiv 0
\end{equation}

This is an example of an{\it \ homogeneous degenerate scaling representation}%
 These representations are inadmissible for the generalized canonical
ensemble. The above considerations give a criterium of\ an homogeneous
nondegenerated scaling representation, the non trivial vanishing of the
curvature tensor determinant:

\begin{equation}
\mathrel{\mathop{\lim }\limits_{N\rightarrow \infty }}%
\det \left( N^{\tau }\partial _{\mu }\partial _{\nu }S_{q}\right) \neq 0
\end{equation}

\subsection{\ The bordering systems.}

The previous analysis showed that the possible application of the formalism
of Tsallis can be found for the bordering systems. An exponential growth of
the density of accessible states during the ThL indicates that this system
will be strongly chaotic due the spectacular quantity of states that it will
be able to access during its evolution in the time. However, using the
previous argument, the bordering systems will have a weaker chaotic behavior.

The q-generalization of the Boltzmann-Gibbs PDF from the Tsallis Formalism$%
\left[ 17,18\right] $ is given by:

\begin{equation}
\omega _{q}\left( X;\beta ,N,a\right) =\frac{1}{Z_{q}\left( \beta
,N,a\right) }e_{q}\left[ -\beta \cdot I_{N}\left( X;a\right) \right]
\end{equation}
where $Z_{q}\left( \beta ,a,N\right) $ is the q-generalized partition
function:

\begin{equation}
Z_{q}\left( \beta ,N,a\right) =\int e_{q}\left[ -\beta \cdot I_{N}\left(
X;a\right) \right] dX
\end{equation}
For this ensemble the {\it q-generalized Laplace Transformation} will be
give by: 
\begin{equation}
Z_{q}\left( \beta ,N,a\right) =\int e_{q}\left[ -\beta \cdot I\right]
W\left( I,N,a\right) \frac{dI}{\delta I_{o}}
\end{equation}
The Laplace Transformation establishes the connection between the
fundamental potentials of both ensembles, the q-generalized {\it Planck\
potential}:

\begin{equation}
P_{q}\left( \beta ,N,a\right) =-\ln _{q}\left[ Z_{q}\left( \beta ,N,a\right) %
\right]
\end{equation}
and generalized Boltzmann entropy given in Eq.$\left( 5\right) $, by the
expression:

\begin{equation}
e_{q}\left[ -P_{q}\left( \beta ,N,a\right) \right] =\int e_{q}\left[ -\beta
\cdot I\right] e_{q}\left[ S_{B-T}\left( I,N,a\right) \right] \frac{dI}{%
\delta I_{o}}
\end{equation}
The q-logarithmic function satisfy the {\it subadditivity relation}:

\begin{equation}
\ln _{q}\left[ xy\right] =\ln _{q}\left[ x\right] +\ln _{q}\left[ y\right]
+\left( 1-q\right) \ln _{q}\left[ x\right] \ln _{q}\left[ y\right] 
\end{equation}
From it is derived the identity:

\begin{equation}
e_{q}\left[ x\right] e_{q}\left[ y\right] =e_{q}\left[ x+y+\left( 1-q\right)
xy\right] 
\end{equation}
The last identity allows to rewrite Eq.$\left( 31\right) $ as:

\begin{equation}
e_{q}\left[ -P_{q}\left( \beta ,N,a\right) \right] =\int e_{q}\left[ -\beta
\cdot I+S_{B-T}\left( I,N,a\right) -\left( 1-q\right) \left( \beta \cdot
I\right) S_{B-T}\left( I,N,a\right) \right] \frac{dI}{\delta I_{o}}
\end{equation}

In the Tsallis case, the lineal form of the Legendre Transformation {\it is
violated} and therefore, in this case the{\it \ ordinary Legendre Formalism
do not establish the correspondence between the two ensembles}.

The nonlineal \ term in the \ q-exponential \ argument {\it violate too the
\ \ homogeneous \ \ scaling \ of \ all \ macroscopic observables for
arbitrary scaling behavior}. The only possibility to preserve the
homogeneous scaling in the q-exponential argument is that all fundamental
macroscopic observables have the same {\it zero-order divergency behavior. }%
The function $f\left( x\right) $ will have a zero-order divergency behavior\
in the infinite limit if it satisfy the conditions:

\begin{equation}
\left( 
\mathrel{\mathop{\lim }\limits_{x\rightarrow \infty }}%
f(x)=\infty \right) \wedge \left( 
\mathrel{\mathop{\lim }\limits_{x\rightarrow \infty }}%
\frac{f(x)}{x^{\alpha }}=0;\text{ }\forall \text{ }\alpha >0\right) 
\end{equation}
The above conditions are only satisfied by systems with scaling behavior
given in $\left( 2.c\right) .$ This kind of scaling behavior must correspond
to systems with so weak chaotic regimen, systems belonging to the {\it chaos
threshold}. This result have been supported by an entire series of works
that indicate that the formalism of Tsallis should be associated to those
systems with a dynamic behavior located in the {\it edge of the chaos (see
in }$\left[ 19,20\right] $){\it . }

For this case, we have to assume the nonlineal Tsallis generalization of the
Legendre Formalism$\left[ 21,22\right] $ given by:

\begin{equation}
\widetilde{P}_{q}\left( \beta ,N,a\right) =Max\left[ c_{q}\beta \cdot
I-S_{B-T}\left( I,N,a\right) \right]
\end{equation}
where $c_{q}=1+\left( 1-q\right) S_{q}$. We recognized the generalization of
the Legendre Transformation for the {\it normalized q-expectations values}.
The maximization leads to the relation:

\begin{equation}
\beta =\frac{\nabla S_{B-T}}{1+\left( 1-q\right) S_{B-T}}\left( 1-\left(
1-q\right) \beta \cdot I\right)
\end{equation}
using the identity:

\begin{equation}
\nabla S_{B}=\frac{\nabla S_{B-T}}{1+\left( 1-q\right) S_{B-T}}
\end{equation}
where $S_{B}$ is the usual Boltzmann entropy, we can rewrite Eq.$\left(
37\right) $ as:

\begin{equation}
\beta =\nabla S_{B}\left( 1-\left( 1-q\right) \beta \cdot I\right)
\end{equation}
solving this equation by successive interactions we find:

\begin{equation}
\beta =\frac{\nabla S_{B}}{\left( 1+\left( 1-q\right) I\cdot \nabla
S_{B}\right) }
\end{equation}
valid only if the following restriction is hold:

\begin{equation}
\left| \left( 1-q\right) I\cdot \nabla S_{B}\right| <1
\end{equation}

The last condition is a very interesting result because it allows to limit
the values of the entropy index. If we apply arbitrarily this formalism to a
pseudoextensive system then $I\cdot \nabla S_{B}$ will not bound in the ThL
and the condition {\it only will be satisfy if \ }$q\equiv 1$, found again
that the Tsallis statistics only will be valid for the bordering systems.
There are many examples in the literature in which the same have been
applied indiscriminately without matter if the systems are extensive or not:
i.e. {\it ideal gas, black body equilibrium emission, and others}. In some
cases, the authors of such as works have introduced some artificial
modifications to the original Tsallis formalism in order to obtain the same
results of the classical thermodynamics, i.e. the {\it q-dependent Boltzmann
constant (see for example in }$\left[ 23\right] )$. The above results
indicate the non applicability of the Tsallis Statistic for these kind of
systems.

As we see, the Tsallis formalism introduces a correlation to the canonical
intensive parameters of the Boltzmann-Gibbs PDF. However, an important
second condition have to be satisfy for the validity of the same, {\it the
stability of the maximum}.

This condition leads to the q-generalization of the Microcanonical
Thermostatistics of D.H.E.Gross . In this approach, the stability of the
Legendre formalism rest in the concavity of the entropy, the negative
definition of the quadratic forms of the curvature tensor. In the Tsallis
case, the curvature tensor have to be modified by:

\begin{equation}
\left( K_{q}\right) _{\mu \nu }=\frac{1}{1-\left( 1-q\right) \widetilde{P}}%
_{q}\left( \left( 2-q\right) \frac{\partial }{\partial I^{\mu }}\frac{%
\partial }{\partial I^{\nu }}S_{B-T}+\left( 1-q\right) \left( \beta _{\mu }%
\frac{\partial }{\partial I^{\nu }}S_{B-T}+\beta _{\nu }\frac{\partial }{%
\partial I^{\mu }}S_{B-T}\right) \right)
\end{equation}

Using the definition, we can rewrite Eq.$\left( 34\right) $ as:

\begin{equation}
e_{q}\left[ -P_{q}\left( \beta ,N,a\right) \right] \simeq \int e_{q}\left[ -%
\widetilde{P}_{q}\left( \beta ,N,a\right) \right] e_{q}\left[ -\frac{1}{2}%
\left( I-I_{M}\right) ^{\mu }\cdot \left( -K_{q}\right) _{\mu \nu }\cdot
\left( I-I_{M}\right) ^{\nu }\right] \frac{dI}{\delta I_{o}}
\end{equation}
The maximum will be stable if the eingenvalues of the q-curvature tensor are 
{\it negative} and {\it very large}. In this case, in the q-generalized
canonical ensemble there will be {\it small fluctuations of the integrals of
movement around the its q-expectation values}. The integration of Eq.$\left(
43\right) $ yields:

\begin{equation}
e_{q}\left[ -P_{q}\left( \beta ,N,a\right) \right] \simeq e_{q}\left[ -%
\widetilde{P}_{q}\left( \beta ,N,a\right) \right] \frac{1}{\delta I_{o}\det^{%
\frac{1}{2}}\left( -\frac{1-q}{2\pi }\left( K_{q}\right) _{\mu \nu }\right) }%
\frac{\Gamma \left( \frac{2-q}{1-q}\right) }{\Gamma \left( \frac{2-q}{1-q}+%
\frac{1}{2}n\right) }
\end{equation}
Let $K_{q}$ be the quantity:

\begin{equation}
K_{q}^{-1}=\frac{1}{\delta I_{o}\det^{\frac{1}{2}}\left( -\frac{1-q}{2\pi }%
\left( K_{q}\right) _{\mu \nu }\right) }\frac{\Gamma \left( \frac{2-q}{1-q}%
\right) }{\Gamma \left( \frac{2-q}{1-q}+\frac{1}{2}n\right) }
\end{equation}
rewriting Eq.$\left( 44\right) $ again:

\begin{equation}
e_{q}\left[ -P_{q}\left( \beta ,N,a\right) \right] \simeq e_{q}\left[ -%
\widetilde{P}_{q}\left( \beta ,N,a\right) +\ln _{q}\left[ K_{q}^{-1}\right]
-\left( 1-q\right) \ln _{q}\left[ K_{q}^{-1}\right] \widetilde{P}_{q}\left(
\beta ,N,a\right) \right]
\end{equation}
arriving finally to the condition:

\begin{equation}
R\left( q;I,N,a\right) =\left| \frac{\ln _{q}\left[ K_{q}^{-1}\right] }{%
\widetilde{P}_{q}\left( \beta ,N,a\right) }\right| \ll 1
\end{equation}

The last condition could be considered as a {\it optimization problem} since
the entropic index is the only one independent variable in the functional
dependency of the physical quantities. The specific value of $q$ could be
chosen  in order to minimize the function $R\left( q;I,N,a\right) $ for all
the possible values of the integrals of movement. This way, the problem of
the determination of the entropic index could be solved in the frame of the
microcanonical theory without appeal to the computational modelation or the
experiment.

Thus, the q-generalized Planck\ potential could be obtained by mean of the
generalized Legendre transformation:

\begin{equation}
P_{q}\left( \beta ,N,a\right) \simeq c_{q}\beta \cdot I-S_{B-T}\left(
I,N,a\right)
\end{equation}
and the q-generalization of the Boltzmann entropy will be equivalent with
the Tsallis entropy in the ThL:

\begin{equation}
S_{q}=-\sum p_{k}^{q}\ln _{q}p_{k}\simeq S_{B-T}
\end{equation}

If the condition $\left( 41\right) $ or $\left( 47\right) $ are not
satisfied , the Tsallis canonical ensemble will not be able to describe the
system in the ThL, it there will be a catastrophe of the generalized Laplace
Transformation.

\section{Final Remarks}

We have analyzed the conditions to satisfy by the systems in order to in the
thermodynamic limit its study through the microcanonical ensemble would be
equivalent to some generalized canonical ensemble, for the special cases of
the Boltzmann-Gibbs Distributions and its generalization in the Tsallis
statistics. This has been carried out in analogy with the traditional
methodology, starting from the properties of scaling of the fundamental
observables of the system during the thermodynamic limit.

We have proven that the traditional Thermoestadistics is valid for the case
of the pseudo-extensive systems, those with an exponential growth of the
accessible states density in the thermodynamic limit, that which allows to
extend to their application to some non extensive systems with this kind of
behavior. In this analysis we have checked the importance that acquires the
local reparametrization invariance of the space of the integrals of movement
for the microcanonical ensemble during the passage from this description to
the canonical description, with the occurrence of a spontaneous symmetry
breaking by mean of the properties of scaling of the system.

On the other hand, when is carried out this analysis for the case of the
Tsallis generalization of the Boltzmann-Gibbs Distributions, we have shown
that the same only can be valid for one special class of the bordering
systems, those with non exponential growth of the density of accessible
states in the thermodynamic limit and zero-order divergency behavior of the
fundamental macroscopic observables, systems located in the chaos threshold.
In this context we have shown an entire series of results of the Tsallis
formalism that in this approach they appear in a natural way: the
q-expectation values, the generalized Legendre transformations of between
the thermodynamic potentials, as well as the conditions for the validity of
the same one, having {\it a priori } the possibility to estimate the value
of the entropic index , without the necessity of appeal to the computational
modelation or the experiment.

Of this study it is suggested the possible existence of an entire spectrum
of the bordering systems that they are not covered by the Tsallis statistic.
We refer to those with a scaling given by the conditions $\left( 2.b\right) $
and $\left( 2.d\right) $.

We summarized the results of this work in the following table:

\begin{center}
\begin{tabular}{|l|l|l|l|}
\hline
\begin{tabular}{l}
{\bf Generic System} \\ 
\ \ \ \ {\bf \ Name}
\end{tabular}
& 
\begin{tabular}{l}
\ \ \ \ \ \ {\bf Scaling behavior} \\ 
\ \ \ {\bf of the states density}
\end{tabular}
& 
\begin{tabular}{l}
\ \ {\bf \ \ \ \ \ \ Chaotic} \\ 
{\bf \ Dynamical Regimen}
\end{tabular}
& 
\begin{tabular}{l}
\ \ \ {\bf Generalized} \\ 
{\bf Canonical Statistic}
\end{tabular}
\\ \hline
Pseudo-Extensive & \ \ \ \ \ \ \ \ \ Exponential & \ \ strongly chaotic & \
\ \ Boltzmann-Gibbs \\ \hline
Bordering Systems & 
\begin{tabular}{l}
\ \ \ \ \ \ \ \ \ \ \ Potential \\ 
sub-Potential, Logarithmic
\end{tabular}
& 
\begin{tabular}{l}
\ \ \ weakly chaotic \\ 
\ \ chaos threshold
\end{tabular}
& 
\begin{tabular}{l}
\ \ \ \ \ {\it \ Unknown} \ \ \ \ \ \ \ \ \ \  \\ 
\ \ \ \ \ \ \ \ Tsallis
\end{tabular}
\\ \hline
\end{tabular}
\end{center}

{\small Table\#1. Equivalence of the microcanonical ensemble in the
thermodynamic limit with a generalized canonical ensemble, equilibrium
properties and dynamical connections.}

\begin{quote}
\bigskip
\end{quote}

\end{document}